# Defect-Correlated Skyrmions and Controllable Generation in Perpendicularly Magnetized CoFeB Ultrathin Films

*Haihong Yin[1,2‡*], Xiangyu Zheng[1,3,4‡], Junlin Wang[1,4‡], Yu Zhou[5], Balati Kuerbanjiang[3], Guanqi Li[3,4], Xianyang Lu[3,4], Yichuan Wang[3], Jing Wu[3*], Vlado K. Lazarov[3], Richard F. L. Evans[3], Roy W. Chantrell[3], Jianwang Cai[5], Bo Liu[6], Hao Meng[6] and Yongbing Xu[1,4*]*

[1] Laboratory of Spintronics and Nanodevice, Department of Electronic Engineering, University of York, York, YO10 5DD, UK

[2] School of Information Science and Technology, Nantong University, Nantong 226019, China

[3] Department of Physics, University of York, York, YO10 5DD, UK

[4] York-Nanjing International Center of Spintronics, Nanjing University, Nanjing 210093, China

[5] Institute of Physics, Chinese Academy of Sciences, Beijing 100190, China

[6] Key Laboratory of Spintronics Materials, Devices and Systems of Zhejiang Province, Hangzhou 311300, China

‡ These authors contributed equally to this work.

* e-mail: hhyin@ntu.edu.cn; jing.wu@york.ac.uk, yongbing.xu@york.ac.uk

**Abstract**

Skyrmions have attracted significant interest due to their topological spin structures and fascinating physical features. The skyrmion phase arises in materials with Dzyaloshinskii-Moriya (DM) interaction at interfaces or in volume of non-centrosymmetric materials. However, although skyrmions were generated experimentally, one critical intrinsic relationship between fabrication, microstructures, magnetization and the existence of skyrmions remains to be established. Here, two series of CoFeB ultrathin films with controlled atomic scale structures are employed to reveal this relationship. By inverting the growth order, the amount of defects can be artificially tuned, and skyrmions are shown to be preferentially formed at defect sites. The stable region and the density of the skyrmions can be efficiently controlled in the return magnetization loops by utilizing first-order reversal curves to reach various metastable states. These findings establish the general and intrinsic relationship from sample preparation to skyrmion generation, offering an universal method to control skyrmion density.

Magnetic skyrmions are chiral quasiparticles and topologically protected, in which the spins point in all of the directions wrapping a sphere. Because of their topologically nontrivial spin textures, magnetic skyrmions exhibit many fascinating features, including emergent electromagnetic dynamics[1], effective magnetic monopoles[2] and topological Hall effects[3, 4]. Magnetic skyrmions were first experimentally observed in B20 noncentrosymmertic crystals at low temperatures and low fields[5-10] and subsequently detected in ferromagnetic (FM)/heavy-metal (HM) thin films with perpendicular magnetic anisotropy (PMA). Due to the mirror symmetry breaking plus spin orbit coupling (SOC)[11-15], interfacial DM interaction is induced at interfaces, which competes with SOC and dipolar interaction and stabilizes skyrmions even up to the room temperature. To date, in various FM/HM thin films such as Fe/Ni/Cu/Ni/Cu[16], Ir/Co/Pt[15, 17], Pt/Co/Ta[18], Pt/CoFeB/MgO[18], Ta/Pt/Co/MgO$_x$/Ta[19], Ta/CoFeB/TaO$_x$[20], and Ta/CoFeB/MgO[21], Néel-type magnetic skyrmions at room temperature have been observed and prototype devices have been achieved[22]. However, the applied magnetic field to generate the skyrmion was previously empirically determined because of the lack of a universal principle [15, 18, 21, 23], and interestingly, the hysteresis loops in them are so similar that curved edges are essential for skyrmion generation despite of so different samples [17, 18, 21, 22, 24]. Curved edges of hysteresis loops are typical manifestations of multiple intermediate states during magnetization, implying some correlation between the material features such as defects and the skyrmions. Generally, the material features are mostly determined in the sample preparation, thus clarifying the intrinsic relationship between fabrication, microstructures, magnetization, and skyrmions can provide a comprehensive perspective on magnetic skyrmions, which enables to establish guidance from sample preparation to skyrmion generation and control.

Here, substrate/Ta/MgO/CoFeB/Ta ultrathin films with PMA are chosen for investigation (**Figure 1**a) because of the large skyrmion size (~1 μm) at room temperature [3, 20-22], which can be easily monitored utilizing a polar magneto-optical Kerr effect (MOKE) microscope. In the overwhelming majority of CoFeB ultrathin films where skyrmions have been reported [3, 20-22], the same growth order of sputtering Ta on CoFeB was adopted for skyrmion generation. Even in studies reported by Yu et al.[21] and Zázvorka et al.[25], an ultrathin Ta interlayer was purposely introduced between CoFeB and MgO layers, which was interpreted as playing a critical role in tuning the PMA by the weakening the Fe−O and Co−O bonds at the interface. However, another possibility arises from the detail of the preparation process. To data, MgO/CoFeB/Ta multilayers are commonly prepared by the sputtering technique. It is known that neutron, proton or heavy ion irradiation can create a large amount of defects in irradiated metallic materials[26]. In sputtering processes, Ta atoms have relatively high momentum due to large Z, thus it is expected that more defects should be induced at Ta/CoFeB interface by sputtering Ta on the bottom CoFeB/MgO. If the growth order is inverted, the interfacial defect density should be far lower. In this work, two series of MgO/CoFeB/Ta ultrathin films with different growth order are employed to investigate the influence of defect **(see Supplementary Information S1)**. Our aim is to establish the intrinsic relationship between fabrication, microstructures, magnetization, and skyrmions, and then propose an efficient and universal method to generate skyrmions controllably by a magnetic field.

**Figure 1**b shows a Kerr hysteresis loop with symmetric curved edges, indicating a gradual multiple intermediate states during switching. **Figure 1**c shows the domain evolution in substrate/Ta(5)/MgO(3)/CoFeB(1.4)/Ta(5) (in nm) film under an out-of-plane field. The magnetization exhibits a typical nucleation reversal process, including four stages of nucleation (**Figure 1**c-1), expansion from nuclei (**Figure 1**c-2 and c-3), domain expansion (**Figure 1**c-4 and c-5) and reversal of hard entities (**Figure 1**c-6). The nucleation-dominated reversal is attributed to the inhomogeneous film. Local defects give rise to a broad inhomogeneous distribution of energy barriers, thus generating local magnetic entities with varying coercivity. Several bubble-like magnetization configurations firstly emerge at low applied fields (**Figure 1**c-1), nucleated at local entities with low coercivity (called "easy centers" here). A small increase of the applied field accelerates the domain growth, and the magnetization configurations transform to a labyrinth phase (**Figure 1**c-4). Then the domain wall expands with increasing field and the labyrinth phase transforms to stripe domains (**Figure 1**c-5). When the field increases close to the saturation, the stripe domains are reversed gradually, condense at local entities with a high coercivity (called "hard centers" here), and transform to bubble-like magnetization configurations (**Figure 1**c-6). Finally, these magnetic bubbles are thoroughly reversed when the applied field is large enough. In previous works on Ta/CoFeB/metal-oxides thin films [3, 20-22], the interfacial DM interaction and the generation of magnetic skyrmions have been confirmed. Here, the magnitude of the interfacial DM interaction is estimated to be $0.35 \pm 0.02$ mJ/m$^2$ in our sample **(see Supplementary Information S2).** The interfacial DM interaction competes with SOC, induces and stabilizes Néel-type chiral domain walls in the substrate/Ta(5)/MgO(3)/CoFeB(1.4)/Ta(5) film. **Figure 2**a-1 and a-3 show typical diameters of ~800 nm for single magnetic skyrmions, whose out-of-plane moment component is shown in the Kerr signal contour mapping as shown in **Figure 2**a-2 and a-4. Therefore, magnetic bubbles generated at low and high magnetic field (**Figure 1**c-1

and c-6) are expected to be Néel-type skyrmions, and the evolution of spin textures can be described as "skyrmions—multidomains—skyrmions" (**Figure 2**b).

We have further carried out two series of comparative magnetization tests for both substrate/Ta(5)/MgO(3)/CoFeB(t)/Ta(5) and substrate/Ta(5)/CoFeB(t)/MgO(3)/Ta(5) (**Figure 1**d) specimens **(see Supplementary Information S1 and S3)**. Perpendicular substrate/Ta(5)/MgO(3)/CoFeB(t)/Ta(5) samples with different CoFeB thickness show similar nucleation reversal process; however, fast wall-jumping behaviors are observed in the magnetization reversal of substrate/Ta(5)/CoFeB(t)/MgO(3)/Ta(5) specimens (**Figure 1**e and f). It is noted that magnetic skyrmions only emerge in substrate/Ta(5)/MgO(3)/CoFeB(t)/Ta(5) specimens, and no skyrmions can be detected in substrate/Ta(5)/CoFeB(t)/MgO(3)/Ta(5) specimens. The vastly different findings can be interpreted by their distinct material features. Specimens in this work are all prepared by the sputtering technique, in which Ta atom has relative high momentum due to high Z. Thus by sputtering Ta layer on bottom CoFeB/MgO, the Ta atomic bombardment would induce more defects. The corresponding STEM investigation indicates the Ta/CoFeB interface in **Figure 3**a is more indistinct than that in **Figure 3**b, confirming higher intermixing degree at Ta/CoFeB interface of substrate/Ta(5)/MgO(3)/CoFeB(1.4)/Ta(5) **(see Supplementary Information S4)**. Our investigation is consistent with the previous studies which found that the sputtering will increase the interface roughness and form a magnetic dead layer[27, 28]. Meanwhile, it is noted that the CoFeB layer in substrate/Ta(5)/MgO(3)/CoFeB(1.4)/Ta(5) is semi-crystalline with some local crystalline CoFeB grains about 1–2 nm are formed (**Figure 3**a), while the CoFeB layer in substrate/Ta(5)/CoFeB(1.2)/MgO(3)/Ta(5) is amorphous (**Figure 3**b). Therefore, the Ta atomic bombardment and the local nanocrystallization can both induce more structural defects, resulting in an inhomogeneous CoFeB layer. In the overwhelming majority of literature reports, the same sputtering of Ta on CoFeB layer was adopted for skyrmion generation in Ta/CoFeB/metal-oxides ultrathin films [3, 20-22]. Clearly, the substrate/Ta(5)/MgO(3)/CoFeB(t)/Ta(5) samples show a defect-correlated skyrmion generation. The sputtering- and naocrystallization-induced defects give rise to the inhomogeneous distribution of energy barriers. Consequently, magnetic skyrmions favor emerging at either easy centers or hard centers, generating skyrmions with opposite magnetization sign as shown in **Figure 1**c-1 and c-6.

A simulation based on the atomistic spin model was performed to investigate the influence of sputtering-induced defects on the magnetic anisotropy **(see Supplementary Information S5)**. As shown in **Figure 3**c, the sputtering-induced Ta atoms are considered as non-magnetic atoms and can be seen as missing atoms in the CoFeB layer. It is assumed that the defects penetrate into the Ta coated layer to the thickness $t_d$ dependent on the bombardment energy. In our simulation, the CoFeB layer is 1.4 nm thick and two different thickness of the defect layer are considered. The anisotropy values are calculated using the constrained Monte Carlo Method [29]; details are given in **Supplementary Information S5.** From the results in **Figure 3**c, when the number of defect atoms increase, the perpendicular anisotropy for the whole system will also increase. Furthermore, the trend of anisotropy changing depends on the number of non-magnetic atoms. The PMA of CoFeB is provided by the Fe−O and Co−O bonds at the MgO interface [21, 27], which means that the intrinsic contribution to the PMA, arising from the interface atoms, is unaffected by the defect sites. However, due to a reduction in the effective magnetic thickness, as shown in **Supplementary Information S5**, the effective anisotropy can be described approximately by the expression

$$K = K'(1 + \frac{t_d \cdot \varepsilon}{t_f}) \qquad (1)$$

where $t_d$, $t_f$ are the thickness of the defect layer and the film thickness respectively.   is the defect concentration and $K' = 0.5 K_0\, a/t_f$ with $a$ the CoFeB lattice spacing, $K_0$ the anisotropy of the CoFeB layer in contact with the MgO layer, and 0.5 is the coefficient calculated from bcc crystal structure. This expression is plotted as solid lines in **Figure 3**c and gives good agreement with the simulations. Thus, the existence of defects will lead to the nonuniform PMA distribution in the whole sample. A local area having high defect density will be hard to be reversed, thus generating hard nucleation point named "hard centers"; reversely, a local area with low defect density should be easy to be reversed, thus generating easy nucleation point named "easy centers".

To generate skyrmions by a magnetic field, a unidirectional field was commonly applied on CoFeB ultrathin films[15, 18, 21, 23], but its value was just previously empirically adjusted by increasing the field strength gradually until magnetic skyrmions are generated[15, 18, 21, 23]. To date, how to realize a controllable skyrmion generation by a magnetic field, such as tuning the density and determining the stable region, is still a challenge. Skyrmions are metastates and only stable in a certain magnetic region. Because the applied field can trigger the phase transition between skyrmions and other metastates, creating magnetic skyrmions is inevitably to overcome energy barriers existing between different spin textures. That means magnetic skyrmions are stable only when they are trapped in magnetic potential wells with local minimum free energy. According to the simulation results in **Figure 3**, the defects induce nonuniform anisotropy distribution in the system, which enhances the fluctuation of energy barriers, leading to the formation of many local magnetic potential wells. With the first-order reversal curves (FORC) technique[30], the return magnetization provides an unique way to achieve various meta-stable states determined by these local potential wells, which can then be utilized to control the density of the skyrmions. The generation of a FORC is generally preceded by the saturation of a system in a positive out-of-plane applied field initially. The magnetization $M$ is then measured starting from a reversal field $H_r$ back to positive saturation, tracing out a FORC. **Figure 4**a shows a family of FORCs measured at different $H_r$ with equal field spacing of 0.65 Oe. For each FORC with $H_r$ near the coercivity, an unusual valley is formed, showing a delayed magnetization behavior that the domain reversal would continue even though the applied field H is decreasing. The delayed magnetization behavior attributes to the magnetic aftereffect, caused by the thermal activation, and evidenced by a time dependent magnetization measurement **(see Supplementary Information S6)**. When $H_r$ is larger than coercivity and close to the saturation field, the delayed magnetization behavior is weakened gradually, accompanied with a clear left-valley-shift (**Figure 4**a).

As analyzed in **Figure 1**b and c, only a few skyrmions are generated by applying a unidirectional H. Differently, in the return magnetization of FORCs (**Figure 4**a), H is not the only parameter controlling skyrmion generation, but the reversal field $H_r$ also plays a critical role. Here, the slope $\chi^d$ for each FORC is defined as a derivative:

$$\chi^d_{H_r}(H_r, H) = \frac{dM(H_r, H)}{dH}|_{H_r = constant} \qquad (2)$$

With the variation of Hr, $\chi^d$ is scanned in the H-Hr plane, mapping out a contour plot (**Figure 4**b) or a 3D plot (**Figure 4**d). Magnetic skyrmions only emerge in Zones 1, 2 and 3 as shown in **Figure 4**b and d. Here, the skyrmion field stability is evaluated by the slope $\chi^d$, and simultaneously the influence of the magnetic aftereffect is considered. For FORCs in Zone 1 (-4.2 Oe < Hr < -1.6 Oe), several magnetic skyrmions are initially nucleated at easy centers. The slope $\chi^d$ in Zone 1 is approximately equal to zero, indicating a slight fluctuation of M caused by H, thus magnetic skyrmions in this region are field stable. However, due to the influence of the magnetic aftereffect, skyrmions in the left of Zone 1 ($H < 0$ Oe, marked by a red dash circle in **Figure 4**b) will finally transform to a multidomain phase after applying a fixed field ($H < 0$ Oe) and waiting time long enough. Therefore, magnetic skyrmions are generated in the negative region of Zone 1, but the real stable region is the positive of Zone 1. For Zone 2, Hr is close to the saturation field and located at the curve edge of the major hysteresis loop (**Figure 4**c, marked in the red line). The FORC start from an initial high negative Hr (-26.3 Oe < Hr < -21.8 Oe), running a return magnetization. Take the FORC with Hr = -21.8 Oe as an example. Due to high Hr, most area of the film has been reversed except some stripe domains (**Figure 4**e-1). With H decreasing, the stripe domains keep on switching due to the magnetic aftereffect, and then transform to magnetic skyrmions at hard centers (**Figure 4**e-2). Because the film cannot be thoroughly saturated in a return magnetization process and the slope $\chi^d$ is a small value fluctuating slightly, the magnetic skyrmions exhibit high field stability, maintaining their bubble-like shapes in a wide field range. With H further decreasing to -6.2 Oe and then to 0 Oe, the skyrmion size enlarges and part of the new skyrmions emerge because of a slight increment of $\chi^d$ (**Figure 4**e-3 and e-4), attributing to the magnetization relaxation with H. In Zone 2, the fluctuation of the FORC slope $\chi^d$ is weakened with Hr increasing and the FORCs approach to the horizontal gradually. In addition, the delayed magnetization is depressed in high field region, thus the magnetic aftereffect influence is restricted in a narrow range ($\Delta H < 3$ Oe, the yellow region in **Figure 4**c), leading to the elevated field stability. Independent measurement is performed to verify the skyrmion stability. Skyrmions are initially generated at Hr = -23.7 Oe and H = -12.0 Oe, and then H is tuned randomly in the range from -2.0 Oe to -22.0 Oe; no deformation is observed in the process, showing high topologically protected stability. For Zone 3, the skyrmion generation is similar to that in the major hysteresis loop, where only several skyrmions can be generated at hard centers. Because the slope $\chi^d$ in Zone 3 is much larger than that in Zone 1 and 2, the skyrmions are rapidly reversed and the sample is thoroughly saturated. As analyzed above, although magnetic skyrmions can be generated in Zone 1, 2 and 3; however, Zone 2 is the only region to generate magnetic skyrmions with both high density and high stability.

**Figure 5**a shows the skyrmion density for different FORCs at a zero applied field. The skyrmion density is strongly affected by $H_r$, and the highest density is $2 \times 10^4$ $mm^{-2}$ (**Figure 5**b-2) corresponding to the FORC with Hr = -22.4 Oe. Here, an energy potential scheme is proposed to interpret the influence of Hr and why few skyrmions are obtained in the major hysteresis loop (**Figure 5**c). In the major hysteresis loop, few magnetic skyrmions are generated at low H because the quantity of easy centers is too few (**Figure 1**b-1). These skyrmions rapidly transform to multidomain phases with H increasing. Although new nuclei are successively generated in the wall propagation process, they are rapidly submerged in the multidomain states (Scheme I of **Figure 5**c). When H increases close to the saturation field, most of hard centers are switched except several with the very high energy barriers, thus the amount of skyrmions generated at high H is also few (Scheme II of **Figure 5**c). Quite

differently, H is decreasing in the return magnetization of FORCs (Zone 2 in **Figure 2**b), thus large quantities of hard centers are exposed with the Zeeman energy decreasing (Scheme II of **Figure 5**c). Consequently, the stripe domains formed initially (H = Hr) are easily trapped at the exposed hard centers, and finally transform to metastable magnetic skyrmions with opposite chirality. When Hr increases further and is closer to the saturation field, fewer hard centers are exposed, thus the skyrmion density decreases with Hr (**Figure 5**b).

In conclusion, we have elucidated the intrinsic relationship between fabrication, microstructures, magnetization and the skyrmion generation in the CoFeB/Ta ultrathin films. The sputtering-induced defects cause generally the nonuniform PMA distribution, which leads to the formation of local magnetic potential wells at defect sites during the magnetization. With the FORC technique, the return magnetization can trap and stabilize skyrmions in these potential wells, and thus the skyrmion density and the stable region can be efficiently controlled. These results establish a universal guidance from sample preparation to skyrmion generation, offering an general method to control skyrmion density via the return magnetization, which should also apply to other PMA FM/HM systems for the skyrmion based spintronics applications.

**Experimental Section**

Two series of substrate/Ta(5)/MgO(3)/CoFeB(t)/Ta(5) and substrate/Ta(5)/CoFeB(t)/MgO(3)/Ta(5) (layer thickness in nm) films deposited on thermally oxidized Si wafers (100) using magnetron sputtering system at RT with base pressure better than $5 \times 10^{-5}$ Torr. The bottom Ta (5 nm) layer is a seed layer deposited on Si substrate firstly. The Ta capping layer of 5 nm thickness was deposited to protect the MgO layer and against the film oxidation. In the fabrication, metallic layers (Ta) were deposited by dc magnetron sputtering, while MgO and CoFeB layers were deposited by rf magnetron sputtering **(see Supplementary Information S1)**. The scanning transmission electron microscope (STEM) was utilized to investigate local defect formation and different material features in two series of CoFeB ultrathin films. All measurements in this work were performed at room temperature.

## Acknowledgements

This work was supported by the National Basic Research Program of China (2016YFA0300803), National Natural Science Foundation of China (No. 61274102, No. 61427812 and No. 11574137), Jiangsu NSF (BK20140054), "The Six Top Talents" of Jiangsu (Grant No. 2016-XCL-052), Key NSF Program of Jiangsu Provincial Department of Education (Grant No. 15KJA510004), Jiangsu Government Scholarship for Overseas Studies (JS-2016-064), Jiangsu Shuangchuan Programme. RFLE gratefully acknowledges the financial support of the Engineering and Physical Sciences Research Council (Grant No. EPSRC EP/P022006/1). H.Yin, X.Zheng and J.Wang contributed equally to this work.

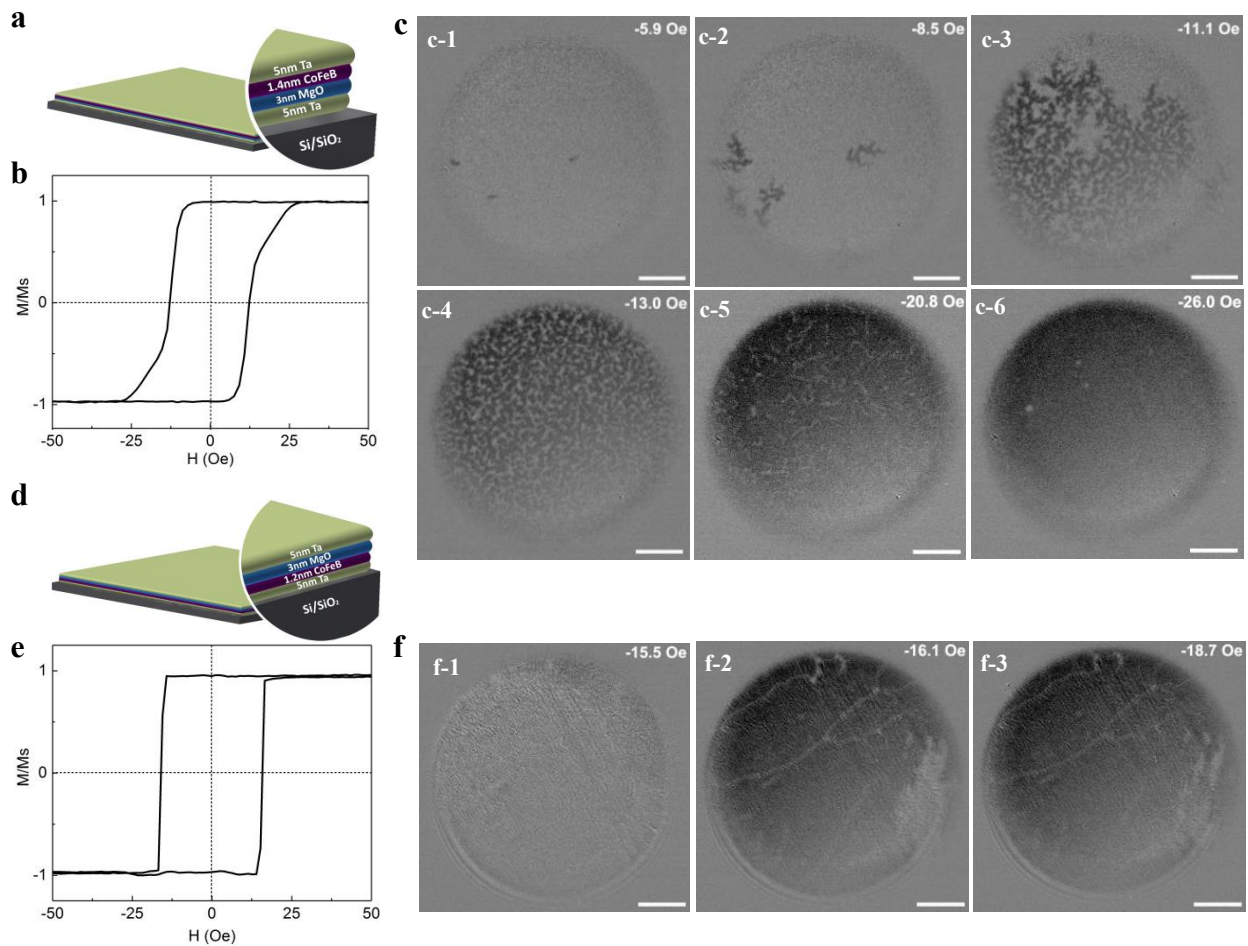

**Figure 1. Sample structure and magnetization reversal. a-c,** Structure schematic (**a**), out-of-plane Kerr hysteresis loop (**b**), and MOKE images (**c**) for substrate/Ta(5)/MgO(3)/CoFeB(1.4)/Ta(5). **d-f**, Structure schematic (**d**), out-of-plane Kerr hysteresis loop (**e**), and MOKE images (**f**) for substrate/Ta(5)/CoFeB(1.2)/MgO(3)/Ta(5). The MOKE images are acquired by changing the out-of-plane field strength after the initial positive field saturation. The scale bar is 20 μm.

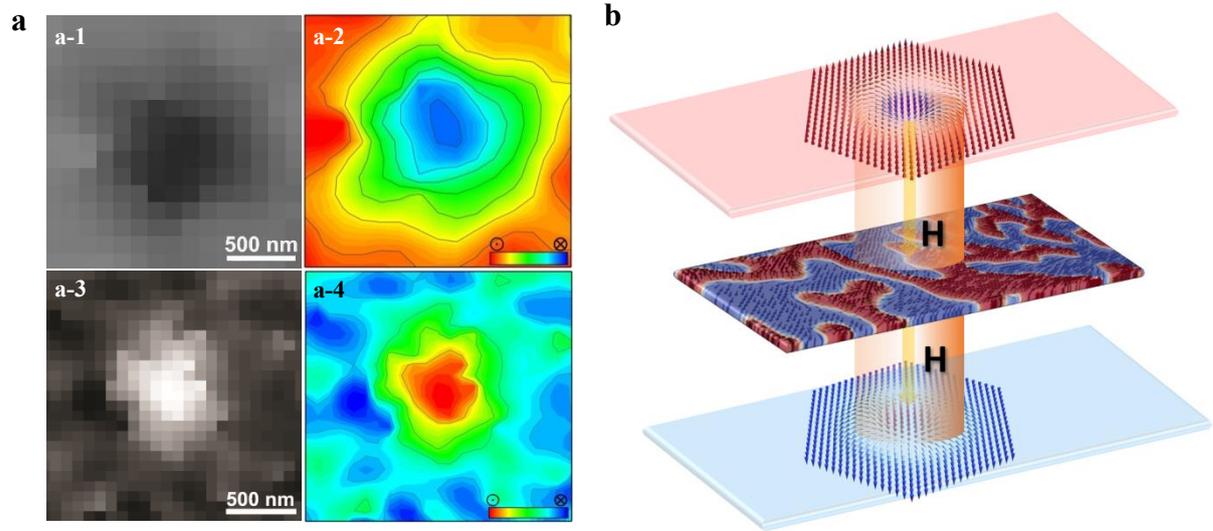

**Figure 2. Analysis and evolution of magnetic skyrmions. a**, Moke images (**a-1** and **a-3**) and the out-of-plane magnetization component (**a-2** and **a-4**) for two magnetic skyrmions with opposite sign. The out-of-plane moment component is plotted by a Kerr signal contour mapping. **b**, Schematic of the evolution between magnetic skyrmions and multidomains in the magnetic reversal process.

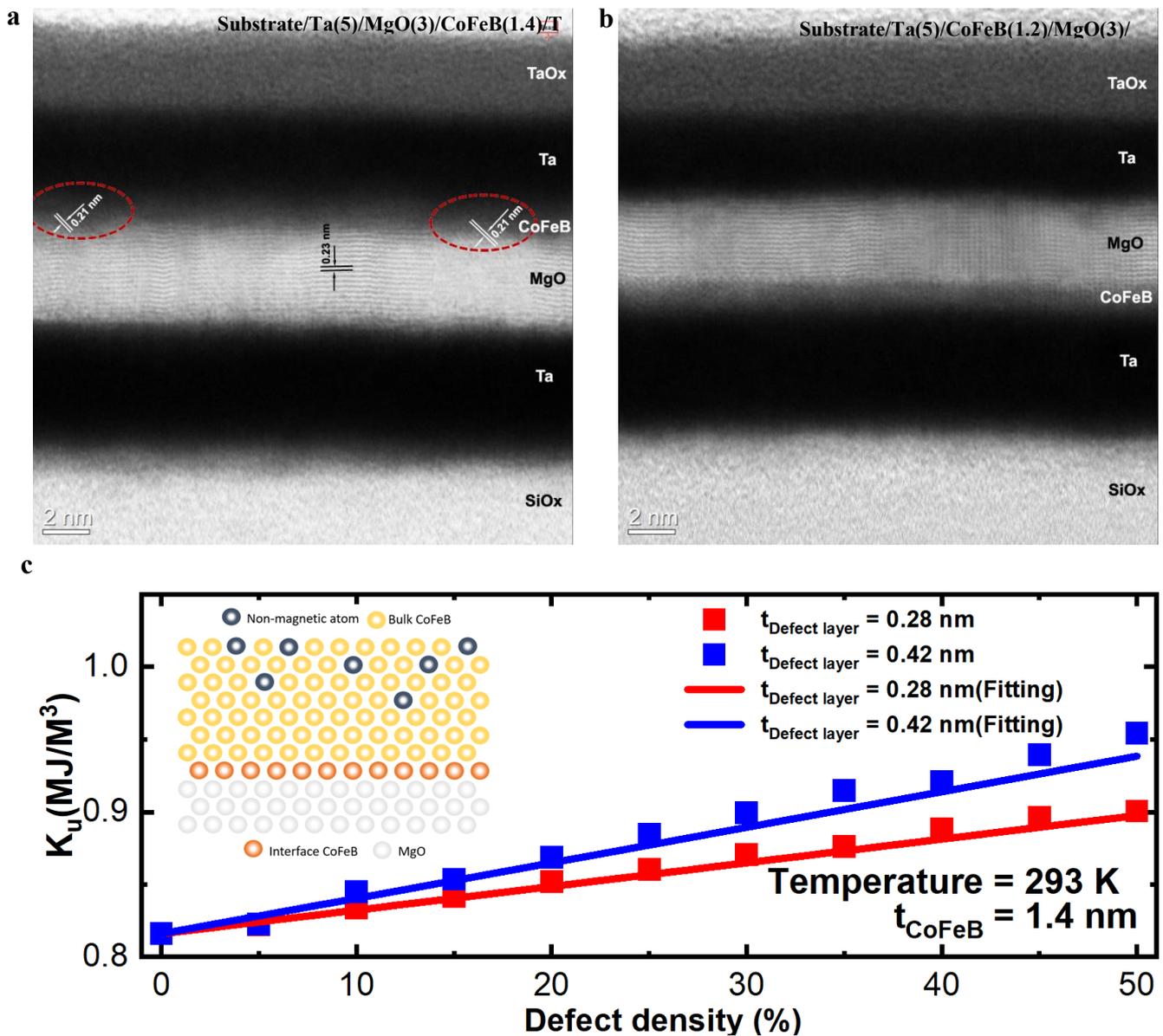

**Figure 3. Sputtering-induced different material features. a-b,** STEM images for substrate/Ta(5)/MgO(3)/CoFeB(1.4)/Ta(5) and substrate/Ta(5)/CoFeB(1.2)/MgO(3)/Ta(5), respectively. The Ta/CoFeB interface in plane **a** is more indistinct than that in plane **b**, and the CoFeB layer in plane **a** is semi-crystalline while the CoFeB layer in plane **b** is amorphous. **c,** The atomistic spin model simulation of defect-influenced magnetic anisotropy energy in the CoFeB single layer. The defect density means the percentage of missing atoms in the defect layer.

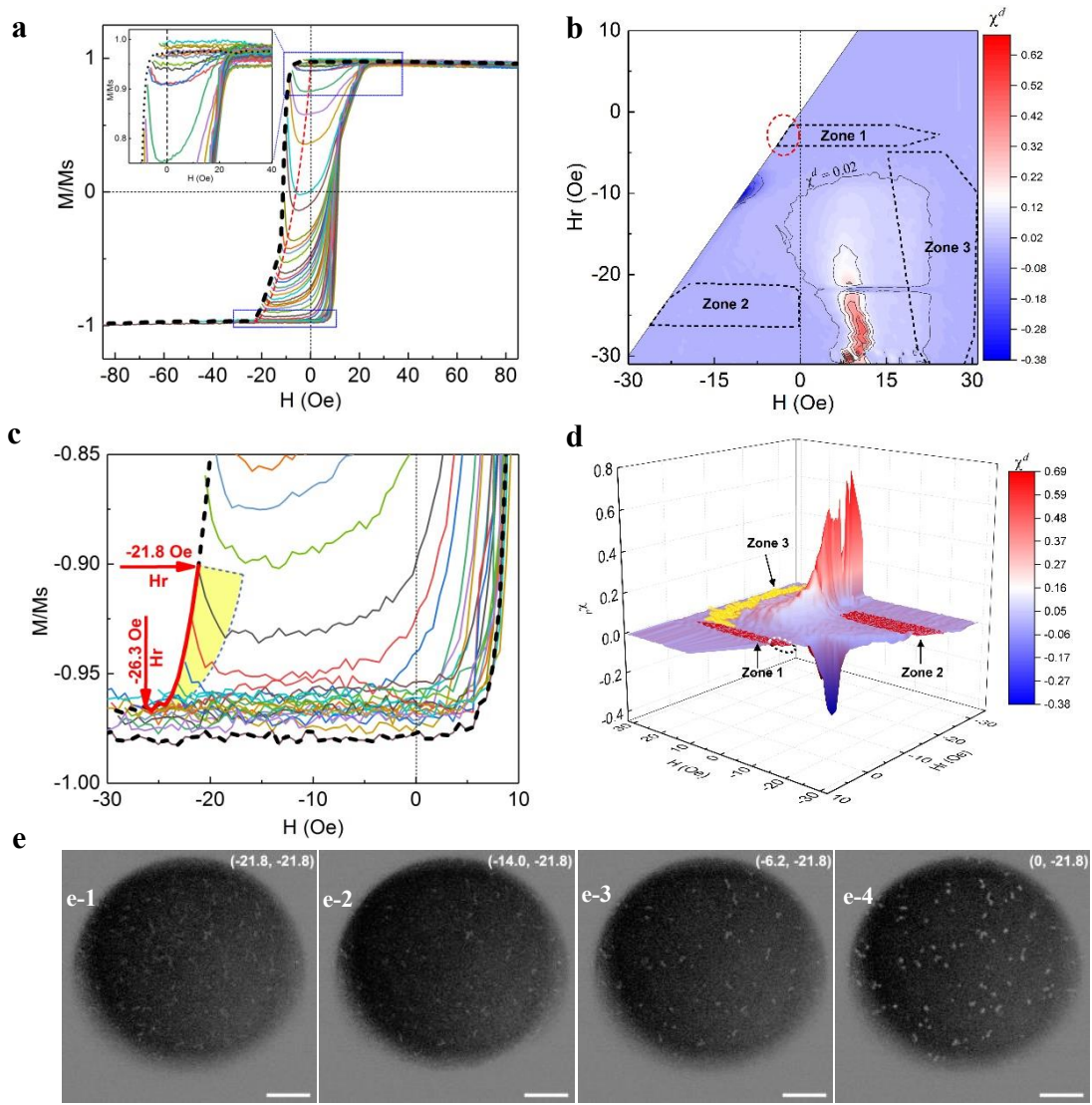

**Figure 4. Return magnetization analysis for substrate/Ta(5)/MgO(3)/CoFeB(1.4)/Ta(5) film. a**, A family of FORCs with an out-of-plane field, showing obvious left-shift valleys (the red dash line). The inset is the enlarged plot for FORCs marked in the right rectangle. **b** and **d**, A contour and a 3D plot of the FORC slope $\chi^d$ versus Hr and H. Zone 1, 2 and 3 is the skyrmion phase diagrams determined by (H, Hr). **c,** The enlarged plot for FORCs marked in the bottom rectangle of plane **a**. The red line indicates Hr required for the skyrmion generation. The yellow region highlights the influence of the magnetic aftereffect. **e**, MOKE images for the transition from stripe domain to skyrmion in the FORC with Hr = -21.8 Oe. The top-right labels are denoted as (H, Hr) with the unit of Oe. The scale bar is 20 μm.

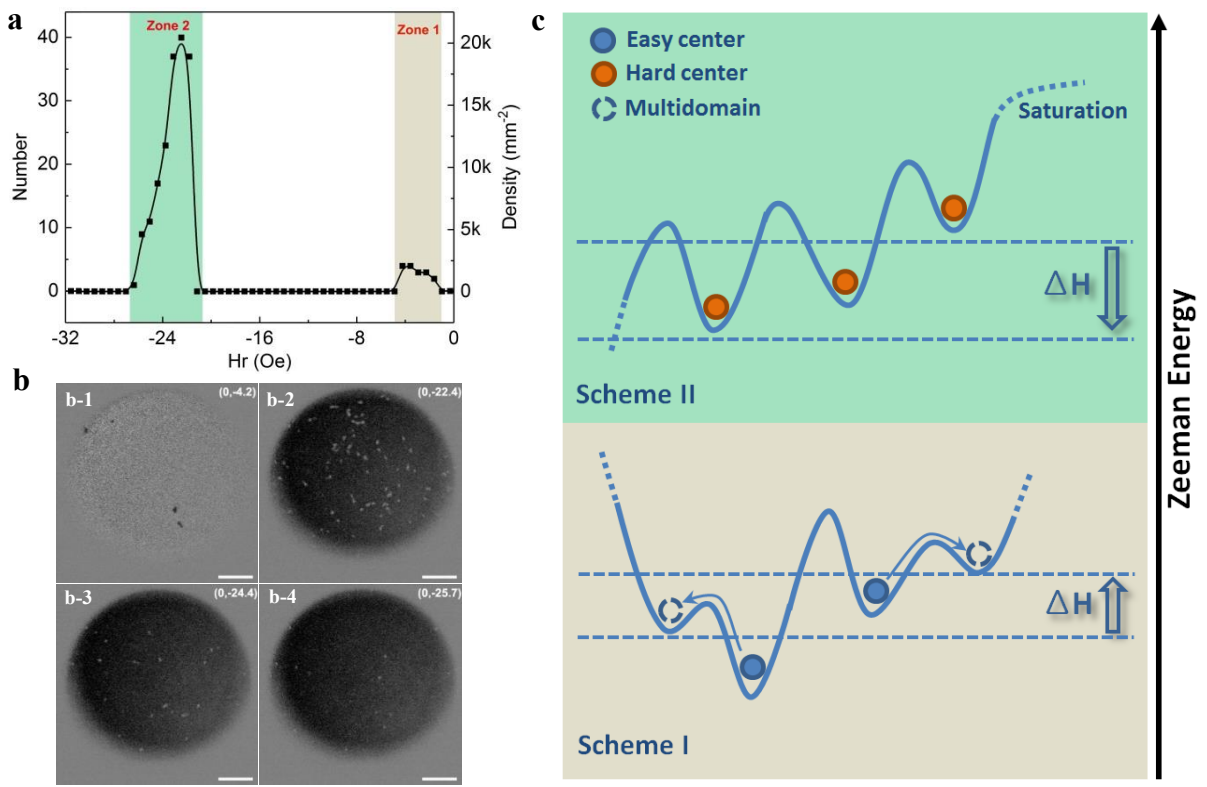

**Figure 5. The density variation and the scheme for skyrmion generation. a-b**, The density/amount and MOKE images of magnetic skyrmions generated at H = 0 Oe for different FORCs. The top-right labels in plane **b** are denoted as (H, Hr) with the unit of Oe. The scale bar is 20 μm. **c**, Magnetic energy potential scheme for skyrmion generation at different Zeeman energy.